\documentclass[11pt,a4paper]{article}
\usepackage{jheppub}
\usepackage[T1]{fontenc}
\usepackage{amsmath}
\usepackage{amssymb}
\usepackage{esint}
\usepackage{amsfonts}
\usepackage{graphicx,color}
\usepackage{slashed}
\usepackage{young}
\usepackage[vcentermath]{youngtab}
\usepackage[utf8]{inputenc}
\usepackage{tensor}
\usepackage{etoolbox}
\usepackage{bbm}

\makeatletter

\newcommand{\be}{\begin{equation}}
\newcommand{\ee}{\end{equation}}
\newcommand{\bea}{\begin{eqnarray}}
\newcommand{\eea}{\end{eqnarray}}

\pdfpageheight\paperheight
\pdfpagewidth\paperwidth

\pdfoutput=1 

    \patchcmd{\maketitle}{\@fpheader}{}{}{}

\title{The Action of the (free) $(4,0)$-theory}
\author[a]{Marc Henneaux,}
\author[a] {Victor Lekeu}
\author[a,b]{and Amaury Leonard}
\affiliation[a]{Universit\'e Libre de Bruxelles and International Solvay Institutes, ULB-Campus Plaine CP231, B-1050 Brussels, Belgium}
\affiliation[b]{Max-Planck-Institut f\"{u}r Gravitationsphysik (Albert-Einstein-Institut),
Am M\"{u}hlenberg 1, \\ DE-14476 Potsdam, Germany}

\abstract
{The $(4,0)$ theory in six dimensions is an exotic theory of supergravity that has been argued to emerge as the strong coupling limit of theories having $N=8$ supergravity as their low energy effective theory in five spacetime dimensions. It has maximal supersymmetry and is superconformal.  Very little is known about this intriguing theory.   While the spectrum of fields occurring in its description has been given and their equations of motion in the absence of interactions have been written down,  no action principle has been formulated, even in the free case.    We extend here previous analyses by writing explicitly the action of the free $(4,0)$ theory from which the equations of motion derive.  The variables of the variational principle are prepotentials adapted to the self-duality properties of the fields. The ``exotic gravitini'', described by chiral fermionic two-forms, are given special attention.  The supersymmetry transformations are written down and   the invariance of the action is explicitly proven.  Even though the action is not manifestly covariant, the symmetry transformations are shown to close according to the $(4,0)$-extended Poincar\'e supersymmetry algebra.   We also discuss exotic supergravity models with fewer supersymmetries. Remarks on dimensional reduction close the paper.}

\makeatother

\begin{document}
\maketitle \flushbottom

\section{Introduction}
\setcounter{equation}{0}

There exist 3 maximal supersymmetry algebras in 6 spacetime dimensions \cite{Strathdee:1986jr}: the $(4,0)$ superalgebra, the $(3,1)$ superalgebra and the $(2,2)$ superalgebra -- as well, of course, as the chirality-reversed superalgebras $(0,4)$ and $(1,3$).  In $(p,q)$, the number $p$ (respectively $q$) refers to the number of spinor charges transforming in the positive (respectively, negative) chiral spinor representation of the Lorentz group  in 6 dimensions (or rather its double cover $Spin(5,1) \simeq SL(2, \mathbb{H})$).  The three maximal superalgebras have $32$ supersymmetries and all reduce to the familiar $N=8$ supersymmetry algebra in 4 spacetime dimensions.  

The non-chiral $(2,2)$ superalgebra is the one that occurs in the dimensional reduction of eleven-dimensional supergravity to six dimensions.  The $R$-symmetry is in that case $usp(4) \oplus usp(4)$ and the supersymmetry charges transform as
$ Q_{\frac12}  \sim (2,1; 4,1) \oplus (1,2; 1,4)$, where  the first two indices refer to the representation of the little algebra 
$so(4) \simeq su(2) \oplus su(2)$ and the last two numbers refer to the representation of the $R$-symmetry.  The graviton supermultiplet is given in the bosonic sector by the representations
$ (3,3;1,1) \oplus (1,3; 5,1) \oplus (3,1; 1,5) \oplus (1,1;5,5) \oplus (2,2; 4,4)$
(one symmetric tensor, five non-chiral two-forms, twenty-five scalars and  sixteen vectors),  making a total of 128 physical degrees of freedom.

The $(4,0)$ and $(3,1)$ superalgebras are less familiar.  We concentrate here on the $(4,0)$-superalgebra.  The $(3,1)$-theory will be discussed in a separate publication \cite{Future}.   

The $(4,0)$ theory has been argued by Hull  \cite{Hull:2000zn,Hull:2000rr,Hull:2001iu,Hull:2000cf} to be the strong coupling limit of theories having $N=8$ supergravity as their low energy effective theory in five spacetime dimensions. The $R$-symmetry algebra in the $(4,0)$-case is $usp(8)$. The bosonic field content of the $(4,0)$ theory is given by
\be (5,1;1) \oplus (3,1; 27)  \oplus (1,1;42) \ee
i.e., a chiral tensor of mixed Young symmetry {\tiny $\yng(2,2)$} (``exotic graviton''),  27 chiral two-forms and 42 scalars, making 128 physical degrees of freedom as in the $(2,2)$-theory.  The fermionic field content is given by 8 fermionic chiral $2$-forms (``exotic gravitini'') and  48 spin-$1/2$ fields,
\be
(4,1; 8) \oplus (2,1; 48)
\ee
which matches the number 128 of bosonic degrees of freedom.  The fields fit into a unitary supermultiplet  of the six-dimensional superconformal group $OSp(8^* \vert 8)$ \cite{Hull:2000zn,Chiodaroli:2011pp}.
The theory is expected to have $E_{6,6}$-symmetry (like maximal supergravity in 5 dimensions), the chiral $2$-forms being in the  $\mathbf{27}$ and the scalars parametrizing the coset $E_{6,6}/USp(8)$, which has dimension $78-36 = 42$.   It can be viewed as the square of the $(2,0)$ theory \cite{Borsten:2017jpt}.  

Now, the discussion of \cite{Hull:2000zn,Hull:2000rr,Hull:2001iu} was performed at the level of the equations of motion.  Although one may develop quantization methods that bypass the Lagrangian, the investigation of the quantum properties lies definitely on more familiar grounds when a self-contained action principle does exist. The central result of our paper is to provide such a local variational principle for the free $(4,0)$-theory.

The construction of an action principle is somewhat intrincate because the theory contains fields subject to self-duality conditions. A typical example of such fields is given by chiral $p$-forms, the curvature of which is self-dual.   While the equations of motion of these fields are manifestly covariant, the action principle from which the equations derive \cite{Henneaux:1988gg} is covariant, but not manifestly so\footnote{One can preserve manifest covariance of the action by introducing auxiliary variables that enter non polynomially in the action (even in the free case)\cite{PST1,PST2,BBS}, but we shall not follow this approach here.}.  The Lagrangian is linear in the time derivatives of the fields.  The situation is reminiscent of the Hamiltonian action principle for a relativistic theory, which is also covariant but not manifestly so.  

 The $(4,0)$-theory contains  chiral $2$-forms and so its action principle will not be manifestly covariant.  But it contains also more exotic chiral tensor fields with mixed Young symmetry, as well as fermionic chiral $2$-forms.  These fields add further difficulties.   In order to implement the chirality condition in the action principle, one must introduce ``prepotentials'' for the chiral tensor fields, as it was done initially for formulating four-dimensional linearized gravity in a manifestly duality-symmetric way \cite{Henneaux:2004jw}.   
 
 The  exotic graviton  ({\tiny $\yng(2,2)$}-tensor) was treated in \cite{Henneaux:2016opm}, where an action principle leading to the chirality equation was explicitly given.  To keep supersymmetry manifest, one must also introduce prepotentials for the exotic gravitini.  The prepotential for the exotic graviton transforms into these prepotentials under supersymmetry, as found in   \cite{Bunster:2012jp,Bunster:2014fca} in the manifestly duality-invariant descriptions of the four-dimensional $(\frac32,2)$ and $(2, \frac52)$ supermultiplets.
 
 A crucial feature of the prepotential formulation is the appearance of the Weyl gauge symmetry: the prepotentials are invariant not only under standard higher-spin diffeomorphisms, but also under higher-spin Weyl transformations.  Furthermore, the number of spatial dimensions is always ``critical''  in the sense that the Weyl tensor,  i.e., the traceless part of the curvature tensor, identically vanishes,  so that one must use analogs of the Cotton tensor to control the Weyl symmetry of the prepotentials \cite{Bunster:2012km,Henneaux:2015cda,Henneaux:2016zlu,Henneaux:2016opm}.  While the presence of the higher-spin Weyl gauge symmetry is still somewhat intriguing, it provides a  very powerful tool to constrain the form of the action.

Our paper is organized as follows.  In the next section (Section \ref{sec:gravitini}), 
we consider the description of the exotic gravitino in six spacetime dimensions. We show that the equations of motion can be rewritten as self-duality conditions completed by a constraint on the fermionic field, which can be solved through the introduction of prepotentials.
In Section \ref{sec:action}, we give explicitly the action of the $(4,0)$-theory.  The action has manifest $USp(8)$-symmetry but is not manifestly covariant.    We then discuss explicitly supersymmetry in Section \ref{sec:susy}.  We give the transformation rules and verify that the action is invariant. We also compute the algebra of the supersymmetry transformations and obtain the super-Poincar\'e algebra.   In Section \ref{sec:lower}, we  provide the actions for the exotic supergravity theories with lower supersymmetry ($(1,0)$, $(2,0)$ and $(3,0)$).  In section \ref{sec:DimRed}, we discuss the dimensional reduction to five dimensions and show that the action of the $(4,0)$-theory reduces to the action of linearized maximal supergravity.  Section \ref{sec:Conclu} contains brief conclusions and comments.  Three appendices complete our paper: Appendix  \ref{app:Gamma6} quickly recalls the central properties of $\Gamma$-matrices in six spacetime dimensions, Appendix \ref{app:usp8} gives our $USp(8)$ conventions, while Appendix \ref{app:cotton5} develops the conformal geometry of a spinorial two-form in five dimensions -- the number of spatial dimensions in six spacetime dimensions.   

Spacetime is flat Minkowski spacetime $\mathbb{R}^{1,5}$ throughout with a Lorentzian metric of ``mostly plus'' signature.

\section{The exotic gravitino}
\label{sec:gravitini}
\setcounter{equation}{0}
\subsection{Fermionic two-form}

The action for a fermionic two-form $\Psi_{\sigma\tau}$ is given by a straightforward generalization of the Rarita-Schwinger action, 
\begin{equation}
S = \int \!d^6\!x\, \bar{\Psi}_{\mu\nu} \Gamma^{\mu\nu\rho\sigma\tau} \partial_\rho \Psi_{\sigma\tau} \label{RS2F6}
\end{equation}
(with $\bar{\Psi}_{\mu\nu} \equiv i \Psi_{\mu\nu}^\dagger \Gamma^0$), where we have taken the spacetime dimension to be six from the outset although the action (\ref{RS2F6}) makes sense in any number of spacetime dimensions. The action is invariant under the gauge transformations
\begin{equation}
\delta \Psi_{\mu\nu} = \partial_\mu \lambda_\nu - \partial_\nu \lambda_\mu = 2 \partial_{[\mu} \lambda_{\nu]} .
\end{equation}
The invariant field strength for this gauge transformation is
\begin{equation}
H_{\mu\nu\rho} = \partial_\mu \Psi_{\nu\rho} + \partial_\nu \Psi_{\rho\mu} + \partial_\rho \Psi_{\mu\nu} = 3 \partial_{[\mu} \Psi_{\nu\rho]} .
\end{equation}
The equation of motion is the generalized Rarita-Schwinger equation
\begin{equation} \label{eom2form}
\Gamma^{\mu\nu\alpha\beta\gamma} H_{\alpha\beta\gamma} = 0 .
\end{equation}
In six spacetime dimensions, one can impose a positive or negative chirality condition (Weyl spinors)  and we shall assume from now on that 
\be
\Gamma_7 \Psi_{\lambda \mu} = \Psi_{\lambda \mu} \label{eq:ChiralPsi}
\ee
(our conventions on $\Gamma$-matrices in six spacetime dimensions are collected in Appendix \ref{app:Gamma6}).  This implies that 
\be
\Gamma_7 H_{\mu\nu\rho} = H_{\mu\nu\rho}
\ee
and that the gauge parameter $\lambda_\nu$ must also be taken to have positive chirality, 
\be
\Gamma_7 \lambda_\nu = \lambda_\nu.
\ee

The equations of motion (\ref{eom2form}) can be split into space and time as follows,
\begin{equation}
\Gamma^{iabc} H_{abc} = 0, \quad \Gamma^{ijabc} H_{abc} + 3 \Gamma^0 \Gamma^{ijab} H_{0ab} = 0 . \label{eq:eomBis}
\end{equation}
The first equation is a constraint on $\Psi_{mn}$ and its spatial gradients, and the second involves the time derivatives of $\Psi_{mn}$ and is dynamical.  They are of first order.  In fact, the action itself is already in Hamiltonian form and can be decomposed as
\begin{equation} \label{eq:2formactionham}
S = \int \!dt \,d^5\!x\, \left( \eta^{ij} \dot{\Psi}_{ij} - \mathcal{H} + \Psi_{0i}^\dagger \mathcal{C}^i + \mathcal{C}^{i\dagger} \Psi_{0i} \right)
\end{equation}
where the conjugate momentum, the Hamiltonian density and the constraint are
\begin{align}
\eta^{ij} &= -i \Psi^\dagger_{kl} \Gamma^{klij} \\
\mathcal{H} &= -\bar{\Psi}_{ij} \Gamma^{ijklm} \partial_k \Psi_{lm} \\
\mathcal{C}^i &= 2i \Gamma^{ijkl} \partial_j \Psi_{kl} .
\end{align}
The momentum conjugate to $\Psi_{0i}$ identically vanishes, and $\Psi_{0i}$ only appears in the action as a Lagrange multiplier for the constraint $\mathcal{C}^i = 0$, which is of course just $\Gamma^{iabc} H_{abc} = 0$.

\subsection{Symplectic Majorana condition}
The spinorial two-form $\Psi_{\lambda \mu}$ is subject to the chirality condition (\ref{eq:ChiralPsi}) but to no reality condition since the Majorana condition on a single chiral spinor cannot be imposed in six spacetime dimensions.  This means, in particular, that $\Psi_{\lambda \mu}$ and $\Psi_{\lambda \mu}^\dagger$ are independent variables in the variational principle. 

It is sometimes convenient to introduce a second chiral spinorial two-form $\Psi_{\lambda \mu}^2$ related to the original spinorial two-form  $\Psi_{\lambda \mu}^1 \equiv \Psi_{\lambda \mu}$ through 
\be
\Psi_{2 \, \lambda \mu}^* \equiv \left(\Psi_{\lambda \mu}^2\right)^* = \epsilon_{21} \mathcal{B} \Psi_{\lambda \mu}^1 = - \mathcal{B} \Psi_{\lambda \mu}^1 \label{eq:usp(2)0}
\ee
where the matrix $\mathcal{B}$ realizes the equivalence between $\left(\Gamma^\mu \right)^*$ and $\Gamma^\mu$ (see Appendix \ref{app:Gamma6}), and where $\epsilon_{ab}$ ($a, b = 1,2$) is the Levi-Civita symbol in the two-dimensional internal space of the $\Psi_{\lambda \mu}^a$'s.  It follows from (\ref{eq:usp(2)0}) that
\be
\Psi_{a \, \lambda \mu}^* \equiv \left(\Psi_{\lambda \mu}^a\right)^* = \epsilon_{ab} \mathcal{B} \Psi_{\lambda \mu}^b  \label{eq:usp(2)1}
\ee
and that the action  (\ref{RS2F6})  can be rewritten as
\begin{equation}
S = \frac{1}{2} \int \!d^6\!x\, \bar{\Psi}_{a \, \mu\nu} \Gamma^{\mu\nu\rho\sigma\tau} \partial_\rho \Psi_{\sigma\tau}^a, \label{ActionUSP2}
\end{equation}
(with $\bar{\Psi}_{a \, \mu\nu} \equiv i \left(\Psi_{ \mu\nu}^a\right)^\dagger \Gamma^0$), exhibiting a manifest $USp(2)$ symmetry under which the spinorial two-forms transform in the ${\mathbf 2}$.  The internal index is moved down when taking complex conjugates because the ${\mathbf 2}^*$ is equivalent to the representation contragredient to the ${\mathbf 2}$ (see Appendix \ref{app:usp8}). 

The condition (\ref{eq:usp(2)1}) is called the ``symplectic Majorana'' condition and trades $\Psi_{\lambda \mu}^*$ for $\Psi_{\lambda \mu}^2$.  The discussion of the prepotentials can be made either in the original single spinor formulation or in the doubled manifestly $USp(2)$ invariant formulation.  For simplicity of notations, we shall carry the construction in the single spinor formulation.  This leads to a single prepotential, which must be doubled by the ``symplectic Majorana rules'' in order to get the prepotentials of the manifestly $USp(2)$ invariant formulation.

The gauge freedom of the theory enables one to impose the light cone gauge where the fields are transverse and obey the necessary $\gamma$-trace conditions.   The physical helicities transform in the $(4,1)$ of the little group algebra $su(2) \oplus su(2)$, motivating the name of ``exotic gravitino'' for the field $\Psi_{\lambda \mu}$.  By contrast, one has the $(3,2)$ for the ``ordinary gravitino'' $\Psi_\lambda$.  In both cases, the representation of the R-symmetry $USp(2)$ is the ${\mathbf 2}$.

\subsection{Self-duality condition}

The equations of motion for the chiral spinorial two-form are equivalent  to the set formed by the self-duality condition on its gauge-invariant curvature $H$ and the constraint,
\begin{equation} \label{twisted2form}
H =  \star H  , \quad  \Gamma^{iabc} H_{abc} = 0 \, .
\end{equation}
The goal of this subsection is to establish this fact.

The  self-duality condition is consistent because $(\star)^2 = 1$, just as for ordinary two-forms in six spacetime dimensions.  If the spinorial two-form were not chiral, the self-duality condition would involve $\Gamma_7$ and would read
\be
 H = \Gamma_7 \star H,
 \ee
 involving both self-dual and anti-self-dual components, which are separated by diagonalizing $\Gamma_7$.
 
In components, the self-duality condition reads
\begin{equation}
H_{0ij} = \frac{1}{3!} \varepsilon_{ijabc}  H^{abc} \; \Leftrightarrow\; H_{abc} = \frac{1}{2} \varepsilon_{abcij}  H\indices{_0^{ij}} .
\end{equation}
The proof of the equivalence of \eqref{eom2form} and \eqref{twisted2form} goes as follows.
\begin{itemize}
\item \eqref{eom2form} $\Rightarrow$ \eqref{twisted2form}: Contracting the dynamical equation with $\Gamma_i$, the constraint implies also
\begin{equation}
\Gamma^{ijk} H_{0jk} = 0, \quad \Gamma^{jk} H_{0jk} = 0 .
\end{equation}
(The second equation follows from the first by contracting with $\Gamma_i$.) Then, contracting the standard identities on the $\Gamma$-matrices
$\Gamma^i \Gamma^{jab} = \Gamma^{ijab} + 3 \delta^{i[j} \Gamma^{ab]}$ and 
$\Gamma^{ij} \Gamma^{ab} = \Gamma^{ijab} + 4 \delta^{[i[a} \Gamma^{b]j]} -2 \delta^{[i[a} \delta^{b]j]} $
with $H_{0ab}$, we get
\begin{align}
\Gamma^{ijab} H_{0ab} = 2 \delta^{a[i} \Gamma^{j]b} H_{0ab}, \quad \Gamma^{ijab} H_{0ab} = 4 \delta^{a[i} \Gamma^{j]b} H_{0ab} +2 H\indices{_0^{ij}}
\end{align}
which together imply
\begin{equation}
\Gamma^{ijab} H_{0ab} = -2 H\indices{_0^{ij}} .
\end{equation}
With this relation, the dynamical equation becomes
\begin{equation}
\Gamma^{ijabc} H_{abc} - 6 \Gamma^0 H\indices{_0^{ij}} = 0 .
\end{equation}
Now, using the identity $\Gamma^{ijabc} = - \varepsilon^{ijabc} \Gamma_0 \Gamma_7$,  multiplying by $\Gamma_0$ and using the chirality condition, we derive
\begin{equation}
\varepsilon^{ijabc}  H_{abc} - 6 H\indices{_0^{ij}} = 0,
\end{equation}
which is exactly the self-duality condition in components.
\item \eqref{twisted2form} $\Rightarrow$ \eqref{eom2form}: Contracting the $abc$ component of the self-duality equation with $\Gamma^{iabc}$ and using the constraint, we get
\begin{equation}
\Gamma^i H_{0ij} = 0, \quad \Gamma^{ij} H_{0ij} = 0 .
\end{equation}
(The second one follows from the first upon contraction with $\Gamma^j$.)
We now use the identity
$\Gamma^{ija} \Gamma^b = \Gamma^{ijab} + 3 \Gamma^{[ij} \delta^{a]b}$,
along with  $\Gamma^{ij} \Gamma^{ab} = \Gamma^{ijab} + 4 \delta^{[i[a} \Gamma^{b]j]} -2 \delta^{[i[a} \delta^{b]j]}$. Contracting them with $H_{0ab}$, we find again $\Gamma^{ijab} H_{0ab} = -2 H\indices{_0^{ij}}$. Using this in the self-duality equation, we obtain the dynamical equation of motion for $\Psi_{\lambda \mu}$ (second of equation (\ref{eq:eomBis})).
\end{itemize}

We thus conclude that the spinorial two-form in six dimensions possesses the remarkable property that its field strength is self-dual, just as the field strengths of its {\tiny $\yng(2,2)$}- and {\tiny $\yng(1,1)$}-bosonic partners of the $(4,0)$ theory.  The self-duality condition does not provide a complete description in the fermionic case, however, since it must be supplemented by the condition $\Gamma^{iabc} H_{abc} = 0$.

\subsection{Prepotential}

The Lagrangian formulations of the bosonic chiral two-form \cite{Henneaux:1988gg} and of the bosonic chiral {\tiny $\yng(2,2)$}-tensor \cite{Henneaux:2016opm} have the following important properties: (i) they involve only spatial tensors, and, in particular, the temporal components of the fields (which are pure gauge) are absent; (ii) the variational equations of motion are not the self-duality conditions ``on the nose''  but an equivalent differential version of them in which the temporal components have been eliminated by taking the appropriate curls\footnote{If one integrates these equations, one gets the self-duality conditions, where the temporal components of the fields re-appear as integration functions. A discussion of the subtleties that arise in the chiral two-form case when the topology of the spatial sections is non-trivial may be found in \cite{Bekaert:1998yp}.}.

A similar formulation for the chiral spinorial two-form exists and is in fact mandatory if one wants to exhibit supersymmetry.  It is obtained by solving the constraint $\Gamma^{iabc} H_{abc} = 0$ in terms of a ``prepotential'', getting rid thereby of its Lagrange multiplier $\Psi_{0 i}$. Once this is done, the only relevant equation that is left is the self-duality condition on the curvature, or rather its differential version
\be
\varepsilon^{rsmij} \partial_m\left(H_{0ij} - \frac{1}{3!} \varepsilon_{ijabc}  H^{abc} \right) = 0 \; \Leftrightarrow\; \varepsilon^{rsmij} \partial_m\left(\partial_0 \Psi_{ij}  \right)  - 2 \, \partial_m H^{mrs}= 0 \label{eq:SDPSI}
\ee
that does not contain $\Psi_{0 i}$.

As shown in Appendix \ref{app:cotton5}, the general solution of the constraint $\Gamma^{iabc} H_{abc} = 0$ can be written in terms of a chiral antisymmetric tensor-spinor $\chi_{ij}$ ($\Gamma_7 \chi_{ij} = \chi_{ij}$, $\chi_{ij} = - \chi_{ji}$) as
\begin{equation}
\Psi^{ab}  = S^{ab}[\chi], \label{eq:RelPsiChi}
\end{equation}
where $S^{ab}[\chi]$ is the Schouten tensor of $\chi_{ij}$,  defined in Subsection \ref{app:subSchouten} as
\begin{equation}
S^{ab}[\chi]= - \left( \delta^{[a}_{[i} \Gamma\indices{^{b]}_{j]}} + \frac{1}{6} \Gamma^{ab} \Gamma_{ij} \right) \varepsilon^{ijklm} \partial_k \chi_{lm} .
\end{equation}
We call $\chi_{ij}$ the prepotential of $\Psi_{ij}$. In other words, the chiral two-form is the Schouten tensor of the prepotential, because of the ``constraint property of the Schouten tensor'' established in \ref{app:subSchouten}.

Furthermore, if $\Psi_{ij}$ is determined up to a gauge transformation $\delta\Psi_{ij} = 2 \partial_{[i} \lambda_{j]}$, then $\chi_{ij}$ is determined up to
\begin{equation}
\delta \chi_{ij} = \partial_{[i} \eta_{j]} + \Gamma_{[i} \rho_{j]}, \label{eq:GaugeChi}
\end{equation}
where $\lambda_i = \Gamma_0 \rho_i$. This is because the Einstein tensor of $\Psi_{ij}$ is equal to the Cotton tensor of the prepotential, and follows then directly from the discussion in the Appendix \ref{app:cotton5}, to which we refer for the details.   The first term in (\ref{eq:GaugeChi}) is the standard gauge transformation of a spinorial two-form, the second term is a generalized Weyl transformation.  When inserted into the Schouten tensor, the first term drops out, while the generalized Weyl transformation induces precisely the gauge transformation of $\Psi_{ij}$.

We now put formula (\ref{eq:RelPsiChi})  in the action \eqref{eq:2formactionham}.   Since the constraint is automatically satisfied, $\mathcal{C}^i = 0$, the Lagrange multipliers $\Psi_{0i}$ disappear. 
The kinetic term becomes
\begin{align}
-i \Psi^\dagger_{ij} \Gamma^{ijkl} \dot{\Psi}_{kl} = -i S_{ij}^\dagger[\chi] \gamma^{ijkl} \dot{S}_{kl}[\chi] = -2 i \chi^\dagger_{ij} \dot{D}^{ij}[\chi]
\end{align}
where $D^{ij}[\chi]$ is the Cotton tensor of $\chi$, defined as
\begin{equation} \label{eq:relDS}
D^{ij}[\chi] = \varepsilon^{ijklm} \partial_k S_{lm}[\chi] .
\end{equation}
The Cotton tensor is invariant under the transformations \eqref{eq:GaugeChi}.
The Hamiltonian is
\begin{align}
\mathcal{H} &= -\bar{\Psi}_{ij} \Gamma^{ijklm} \partial_k \Psi_{lm} =  i \Psi^\dagger_{ij} \varepsilon^{ijklm} \partial_k \Psi_{lm} \\
&= i S^\dagger_{ij}[\chi] D^{ij}[\chi] = -i \chi^\dagger_{ij} \varepsilon^{ijklm} \partial_k D_{lm}[\chi].
\end{align}

The action is therefore
\begin{align}
S[\chi] = -2i \int \!dt \,d^5\!x\, \chi^\dagger_{ij} \left( \dot{D}^{ij}[\chi] - \frac{1}{2} \varepsilon^{ijklm} \partial_k D_{lm}[\chi] \right).
\end{align}
The equations of motion obtained by varying the prepotential are 
\be
\dot{D}^{ij}[\chi] - \frac{1}{2} \varepsilon^{ijklm} \partial_k D_{lm}[\chi] = 0
\ee
which is nothing but the self-duality condition in the form (\ref{eq:SDPSI}), as follows from definitions \eqref{eq:RelPsiChi} and \eqref{eq:relDS}.

\section{The action of the $(4,0)$ theory}
\label{sec:action}
\setcounter{equation}{0}

\subsection{Explicit form}

The action is a sum of five terms, one for each set of fields in the supermultiplet,
\be
S = S_{{\tiny  \yng(2,2)}} +  S_{{\tiny  \yng(1,1)}_F} + S_{{\tiny  \yng(1,1)}_B} + S_{\frac12} + S_{0} .
\ee
We  describe each of these contributions in turn.

\begin{itemize}

\item The exotic graviton

This is a real tensor field of mixed Young symmetry $(2,2)$ with self-dual field strength. It is a singlet under $USp(8)$. The action for such a chiral tensor was derived in \cite{Henneaux:2016opm}.  The variables of the variational principle are the components of the prepotential,  denoted by $Z_{ijkl}$.  The prepotential is real,  $(Z_{ijkl})^* = Z_{ijkl}$. 
The gauge symmetries of the prepotential are generalized diffeomorphisms and generalized Weyl transformations,
\begin{equation}
\delta Z_{ijkl} = \xi_{ij[k,l]} + \xi_{kl[i,j]} + \delta_{[i[k} \lambda_{j]l]} . \label{eq:gaugeZ}
\end{equation}
The action reads
\begin{equation}
S_{{\tiny  \yng(2,2)}} = \frac{1}{2} \int \! d^6\! x \, Z_{mnrs} \left( \dot{D}^{mnrs} - \frac{1}{2} \varepsilon^{mnijk} \partial_k D\indices{_{ij}^{rs}} \right) . \label{eq:actionZ}
\end{equation}
where $D_{ij kl}$ is the Cotton tensor of $Z^{mnrs}$ \cite{Henneaux:2016opm},
\be 
D_{ij kl} = \frac{1}{3!} \varepsilon_{ijabc} \partial^{a} S\indices{^{bc}_{kl}}.
\ee
Here, $S\indices{^{ij}_{kl}} = G\indices{^{ij}_{kl}} - 2 \delta^{[i}_{[k} G\indices{^{j]}_{l]}} + \frac{1}{3} \delta^i_{[k} \delta^j_{l]} G$ is the ``Schouten tensor'' of $Z^{mnrs}$, where $G\indices{^{ij}_{kl}} =  R\indices{^{ij}_{kl}} - 2 \delta^{[i}_{[k} R\indices{^{j]}_{l]}} + \frac{1}{3} \delta^i_{[k} \delta^j_{l]} R$ is its ``Einstein tensor''. The tensor $R\indices{^{ij}_{kl}}$ is the ``Ricci tensor'', i.e., the trace of the Riemann tensor
$ R_{i_1 i_2 i_3 j_1 j_2 j_3} = \partial_{[i_1}Z_{i_2 i_3][ j_1 j_2, j_3]}$, where the comma denotes a derivative (we use a similar notation for higher-order traces).
The Cotton tensor $D_{ij kl}$  is a $(2,2)$-tensor which is invariant under all gauge symmetries (\ref{eq:gaugeZ}), as well as identically transverse and traceless, $\partial_i D^{ij rs} = 0 = D^{ij rs}  \delta_{js}$.  Furthermore, a necessary and sufficient condition for $Z_{ijrs}$ to be pure gauge is that its Cotton tensor vanishes.

The Cotton tensor contains three derivatives of the prepotential and the action (\ref{eq:actionZ}) is therefore of fourth order in derivatives (but of first order in the time derivative).

\item The exotic gravitini

These have been discussed in the previous section. As shown there, the exotic gravitini  are described by fermionic $2$-form prepotentials $\chi^A_{ij}$ which are chiral,
\begin{equation}
\Gamma_7 \chi^A_{ij} = + \chi^A_{ij}.
\end{equation}
The $R$-symmetry is $USp(8)$ and the exotic gravitini transform in the ${\mathbf 8}$.  So there are 8 gravitini, labelled by the $USp(8)$ index $A$ ($A = 1, \cdots, 8$). The reality conditions are given by the symplectic Majorana conditions,
\begin{equation}
 \chi^*_{A\,ij} = \Omega_{AB} \mathcal{B} \chi^B_{ij} .
\end{equation}
These reality conditions are consistent as discussed in Appendix \ref{app:usp8}.

The gauge symmetries are
\begin{equation}
\delta \chi^A_{ij} = 2 \partial_{[i} \eta^A_{j]} + 2 \Gamma_{[i} \rho^A_{j]} .
\end{equation}
and involve both a ``generalized diffeomorphism'' and a ``generalized Weyl transformation''.  The action reads
\begin{equation}
S_{{\tiny  \yng(1,1)}_F} = -2i \int \!d^6\!x\, \chi^{\dagger}_{A\, ij} \left( \dot{D}^{A\,ij} - \frac{1}{2} \varepsilon^{ijklm} \partial_k D^A_{lm} \right)
\end{equation}
where the $D^{A\,ij} \equiv D^{ij}[\chi^A]$ are the Cotton tensors. The action is of third order in derivatives (but again of first order in the time derivative).

\item The chiral $2$-forms

The action for a chiral two-form has been given in \cite{Henneaux:1988gg}.  The chiral two-forms are in the $\mathbf {27}$ of $USp(8)$.  There are thus 27 of them, labelled by the antisymmetric pair $[AB]$ with the constraint
\begin{equation}
A^{AB}_{ij} \Omega_{AB} = 0 . \label{eq:Cons2F}
\end{equation}
The reality condition $(A^{AB}_{ij})^* = A^{AB}_{ij}$ is not compatible with $USp(8)$ covariance, as discussed in Appendix \ref{app:usp8}. Instead, we impose
\begin{equation}\label{eq:Areality}
A^*_{AB\, ij} = \Omega_{AA'} \Omega_{BB'} A^{A'B'}_{ij} .
\end{equation}
The consistency condition $(A^{AB}_{ij})^{**} = A^{AB}_{ij}$ is satisfied because there is an even number of $\Omega$ matrices in equation \eqref{eq:Areality}.

The gauge transformations are
\begin{equation}
\delta A^{AB}_{ij} = 2 \partial_{[i} \xi^{AB}_{j]}
\end{equation}
and the action reads
\begin{equation}
S_{{\tiny  \yng(1,1)}_B} = -\frac{1}{2} \int \! d^6\!x \, A^*_{AB\,ij} \left( \dot{\mathcal{B}}^{AB\,ij} - \frac{1}{2}\varepsilon^{ijklm} \partial_k \mathcal{B}^{AB}_{lm} \right) 
\end{equation}
where the magnetic fields $\mathcal{B}^{AB\, ij}$ are given by
\be
\mathcal{B}^{AB\, ij} = \frac{1}{2} \varepsilon^{ijklm} \partial_k A^{AB}_{lm} .
\ee
The action is of second order in derivatives (and of first order in the time derivative).

It is remarkable that the actions for the exotic graviton, exotic gravitini and chiral $2$-forms have the similar structure ``prepotential $\times$ time derivative of the Cotton tensor''  minus  ``prepotential $\times$ curl of the Cotton tensor'', if one recalls that the chiral $2$-form is its own prepotential so that the magnetic fields can be viewed as the Cotton tensors.  This is the structure characteristic of the descriptions where duality is put in the foreground \cite{Henneaux:2016zlu,Future2}.

\item The spin-$1/2$ fields

These are Dirac fermions $\psi^{ABC}$ subject to the chirality condition
\begin{equation}
\Gamma_7 \psi^{ABC} = + \psi^{ABC}, 
\end{equation}
and transforming in the $\mathbf{48}$ of $USp(8)$.  They are labelled by a completely antisymmetric triplet of indices $[ABC]$ with the constraint
\begin{equation}
\psi^{ABC} \Omega_{AB} = 0 . \label{eq:Cons12F}
\end{equation}
The reality conditions are
\begin{equation}
\psi^*_{ABC} = \Omega_{AA'} \Omega_{BB'} \Omega_{CC'} \mathcal{B} \psi^{A'B'C'} .
\end{equation}
There are no gauge transformations and the action is just a sum of Dirac actions,
\begin{align}
S_{\frac12} &= - \int \! d^6\! x \, \bar{\psi}_{ABC} \Gamma^\mu \partial_\mu \psi^{ABC} \nonumber \\
&= \int \! d^6\! x \,i \psi_{ABC}^{\dagger} \left( \dot{\psi}^{ABC} - \Gamma^0 \Gamma^i \partial_i \psi^{ABC} \right) .
\end{align}
It is of first order in derivatives.

\item The scalar fields

They transform in the $\mathbf{42}$ of $USp(8)$ and are labelled by a completely antisymmetric quadruplet of indices $[ABCD]$ with the constraint
\begin{equation}
\phi^{ABCD} \Omega_{AB} = 0 .\label{eq:ConsScal}
\end{equation}
The reality conditions are
\begin{equation}
\phi^*_{ABCD} = \Omega_{AA'} \Omega_{BB'} \Omega_{CC'} \Omega_{DD'} \phi^{A'B'C'D'} .
\end{equation}
The momenta $\pi^{ABCD}$ satisfy the same conditions.
There are no gauge transformations and the action in hamiltonian form reads
\begin{equation}
S_{0} = \int \! d^6\! x \, \left( \pi^*_{ABCD} \dot{\phi}^{ABCD} - \frac{1}{2} \pi^*_{ABCD} \pi^{ABCD} - \frac{1}{2} \partial_i \phi^*_{ABCD} \partial^i \phi^{ABCD}\right)  .
\end{equation}

\end{itemize}

\subsection{Equations of motion}

The equations of motion following from the action are easily found to be
\be
\dot{D}^{mnrs} = \frac{1}{2} \varepsilon^{mnijk} \partial_k D\indices{_{ij}^{rs}}
\ee
(exotic graviton),
\be \label{eq:eomchi}
\dot{D}^{A\,ij} = \frac{1}{2} \varepsilon^{ijklm} \partial_k D^A_{lm}
\ee
(exotic gravitini),
\be
\dot{\mathcal{B}}^{AB\,ij} = \frac{1}{2}\varepsilon^{ijklm} \partial_k \mathcal{B}^{AB}_{lm}
\ee
(chiral two-forms),
\be
\dot{\psi}^{ABC} = \Gamma^0 \Gamma^i \partial_i \psi^{ABC}
\ee
(spin-$\frac12$ field) and
\be
\dot{\phi}^{ABCD} = \pi^{ABCD} \; , \quad \dot{\pi}^{ABCD} = \partial_i \partial^i \phi^{ABCD}
\ee
(scalars).

The equations of motion of the exotic graviton, exotic gravitini and chiral $2$-forms take a similar form and equate the time derivative of their respective Cotton tensors to their spatial curl.
 
\section{Supersymmetry}
\label{sec:susy}
\setcounter{equation}{0}

Although not manifestly so, the action of the $(4,0)$-theory is Poincar\'e invariant.  This is because each individual piece is Poincar\'e invariant.  This was verified explicitly for the exotic graviton in \cite{Henneaux:2016opm} and for the chiral bosonic two-form in \cite{Henneaux:1988gg}.   The other fields (exotic gravitini, spin-$\frac12$ fields, scalar fields) have a Lorentz-invariant action and Poincar\'e invariance is not an issue for them.

The action is also invariant under $(4,0)$ supersymmetry. We first give the supersymmetry transformations and then check the algebra.

\subsection{Supersymmetry transformations}

The supersymmetry variations only mix fields that have one more or one less $USp(8)$ index, i.e.
\begin{equation}
Z_{ijkl} \,\longleftrightarrow\, \chi^A_{ij} \,\longleftrightarrow\, A^{AB}_{ij} \,\longleftrightarrow\, \psi^{ABC} \,\longleftrightarrow\, \pi^{ABCD},\, \phi^{ABCD} .
\end{equation}
In terms of representations of the little algebra $so(4) \simeq su(2) \oplus su(2)$, this corresponds to fields of neighbouring ``spin'': $(5,1)$, $(4,1)$, $(3,1)$, $(2,1)$ and $(1,1)$.
The canonical dimensions of the various objects (fields and supersymmetry parameter) appearing in the supersymmetry variations are
\begin{center}
\begin{tabular}{c|c|c|c|c|c|c|c}
$Z_{ijkl}$ & $\chi^A_{ij}$ & $A^{AB}_{ij}$ & $\psi^{ABC}$ & $\pi^{ABCD}$ & $\phi^{ABCD}$ & $\epsilon^A$ & $\partial_\mu$ \\ \hline
$1$ & $3/2$ & $2$ & $5/2$ & $3$ & $2$ & $-1/2$ & $1$
\end{tabular}
\end{center}
where $\epsilon^A$ are the supersymmetry parameters and where we have also listed for completeness the canonical dimension of $\partial_\mu$.  We take the supersymmetry parameters to be symplectic Majorana-Weyl spinors of negative chirality,
\begin{equation}
\Gamma_7 \epsilon^A = - \epsilon^A, \quad \epsilon^*_A = \Omega_{AB} \mathcal{B} \epsilon^B ,
\end{equation}
and so $\Gamma_7 \Gamma_0 \epsilon^A = + \Gamma_0 \epsilon^A$.

The variations containing fields with one index more are easy to guess: the $USp(8)$ index on the supersymmetry parameter $\epsilon^A$ must be contracted, and not many possibilities remain with the correct dimension, spatial index structure, $USp(8)$ covariance, chirality, and reality conditions. From those variations, we get the others by requiring the invariance of the kinetic terms and projecting on the appropriate $USp(8)$ representation.
The end result is
\begin{align}
\delta_{\epsilon} Z_{ijkl} &= \alpha_1 \mathbb{P}_{(2,2)} \left( \bar{\epsilon}_A \Gamma_{ij} \chi^A_{kl} \right)  \\
\delta_{\epsilon} \chi^A_{ij} &= - \frac{\alpha_1}{4.3!^3} \partial^r Z\indices{_{ij}^{kl}} \varepsilon_{pqrkl} \Gamma^{pq} \Gamma^0 \epsilon^A + \frac{\alpha_2}{2} A_{ij}^{AB} \Omega_{BC} \Gamma^0 \epsilon^C \\
\delta_{\epsilon} A^{AB}_{ij} &= \alpha_2 \left( 4 \bar{\epsilon}_C S^{[A}_{ij} \Omega^{B]C} + \frac{1}{2} \Omega^{AB} \bar{\epsilon}_{C} S^C_{ij} \right) + \alpha_3 \bar{\epsilon}_{C} \Gamma_{ij} \psi^{ABC} \\
\delta_{\epsilon} \psi^{ABC} &= - \frac{\alpha_3}{2} \Gamma^{ij} \Gamma^0 \left( \mathcal{B}^{[AB}_{ij} \epsilon^{C]} - \frac{1}{3} \Omega^{[AB} \mathcal{B}^{C]D}_{ij} \Omega_{DE} \epsilon^E \right) \\
&\quad + \alpha_4 \left( \pi^{ABCD} \Omega_{DE} \Gamma^0 \epsilon^E + \partial_i \phi^{ABCD} \Omega_{DE} \Gamma^i \epsilon^E \right) \nonumber \\
\delta_{\epsilon} \phi^{ABCD} &= \alpha_4 \left( 2 \bar{\epsilon}_{E} \psi^{[ABC} \Omega^{D]E} + \frac{3}{2} \bar{\epsilon}_E \Omega^{[AB} \psi^{CD]E} \right) \\
\delta_{\epsilon} \pi^{ABCD} &= \alpha_4 \left( 2 \bar{\epsilon}_{E} \Gamma^0 \Gamma^i \partial_i \psi^{[ABC} \Omega^{D]E} + \frac{3}{2} \bar{\epsilon}_E \Gamma^0 \Gamma^i \Omega^{[AB} \partial_i \psi^{CD]E} \right) ,
\end{align}
where $ \mathbb{P}_{(2,2)}$ is the projector on the $(2,2)$-Young symmetry, which takes the explicit form
\begin{equation}
\mathbb{P}_{(2,2)} \left( \bar{\epsilon}_A \Gamma_{ij} \chi^A_{kl} \right) = \frac{1}{3} \left( \bar{\epsilon}_A \Gamma_{ij} \chi^A_{kl} + \bar{\epsilon}_A \Gamma_{kl} \chi^A_{ij} -2 \bar{\epsilon}_A \Gamma_{[i[k} \chi^A_{l]j]} \right)
\end{equation}
in this case, and where $S_{ij}^A$ is the Souten tensor of $\chi^A_{ij}$.  Here,  the matrix $\Omega^{AB}$ (with indices up) is defined through $\Omega^{AB}\Omega_{CB} = \delta^A_C$  and is numerially equal to $ \Omega_{AB}$ (see Appendix \ref{app:Gamma6}).   Having $\epsilon^A$ to be of negative chirality while the fields have positive chirality makes the variations of the bosonic fields not identically zero and gives the correct chirality to those of the fermionic fields. 

These transformations leave not only the kinetic term invariant but one  also verifies that they leave the Hamiltonian invariant.

The real constants $\alpha_1$ to $\alpha_4$ are free at this stage since the action is invariant for any values of them. They will be fixed in Eq. (\ref{eq:RelAlpha}) below (up to an overall factor) through the requirement that the supersymmetry transformations close according to the standard supersymmetry algebra.

For later purposes, it is convenient to compute the supersymmetry transformations of the gauge-invariant tensors.  These are
\begin{align}
\delta_{\epsilon} D_{ijkl}[Z] &= \frac{\alpha_1}{(3!)^3} \mathbb{P}_{(2,2)} \left( \bar{\epsilon}_A \varepsilon_{pqrij} \Gamma^{pq} \partial^r D^A_{kl} \right) \\
\delta_{\epsilon} D_{ij}[\chi^A] &= - \frac{\alpha_1}{4} D_{ijkl}[Z] \Gamma^{kl} \Gamma^0 \epsilon^A \\
&\quad +\frac{\alpha_2}{3} \left( \varepsilon_{ijklm} \partial^k \mathcal{B}^{AB\,lm} + \varepsilon_{ijklm} \partial^k \Gamma\indices{^l_p}\mathcal{B}^{AB\,mp} + 2 \Gamma_k \Gamma_0 \partial^k \mathcal{B}^{AB}_{ij} \right) \Omega_{BC} \Gamma^0 \epsilon^C \nonumber \\
\delta_{\epsilon} \mathcal{B}^{AB}_{ij} &= \alpha_2 \left( 2 \bar{\epsilon}_C D^{[A}_{ij} \Omega^{B]C} + \frac{1}{4} \Omega^{AB} \bar{\epsilon}_{C} D^C_{ij} \right) + \frac{\alpha_3}{2} \bar{\epsilon}_C \varepsilon_{ijklm} \Gamma^{lm} \partial^k \psi^{ABC} .
\end{align}
These variations involve only the gauge invariant objects, as they should.

\subsection{Supersymmetry algebra}

As we mentioned above, the action of the $(4,0)$-theory is Poincar\'e invariant \cite{Henneaux:1988gg,Henneaux:2016opm} although not manifestly so.  The Poincar\'e generators $P_\mu$ and $M_{\mu \nu}$  close therefore according to the Poincar\'e algebra.  

In order to establish the $(4,0)$-supersymmetry algebra, one needs to verify that the anticommutator of two supersymmetries gives a space-time translation, as well as the other commutation relations involving the supercharges.  This is the question on which we focus in this section.

Before proceding with the computation, we stress that in the standard Hamiltonian formalism, the variation  $\epsilon^0 P_0 F$ of a dynamical variable $F$ under a time translation is given by its Poisson bracket $ [F,H]$ with the Hamiltonian times the parameter $\epsilon^0$ of the time translation.  It is a function of the phase space variables only and not of their time derivatives.  When one uses the equations of motion, $\epsilon^0 P_0 F$ becomes of course equal to $\epsilon^0 \partial_0 F$.  The same feature holds for the above ``Hamiltonian-like''  first-order action of the $(4,0)$-theory.
Similarly,  the supersymmetry transformations do not involve time derivatives of the variables.  For this reason, their algebra  cannot contain time derivatives either.  This is again a well known feature of the Hamiltonian formalism,  when transformations are written in terms of phase space variables.  It occurs in the Hamiltonian formulation of  non-exotic supersymmetric theories  as well. To compare with the familiar form of the supersymmetry algebra acting on the fields where time derivatives appear, one must use the equations of motion.  

We first compute the anticommutator of two supersymmetry transformations.  We carry this task for the gauge-invariant curvatures, for which the computation is simpler.  For non gauge-invariant fields,  the commutator of supersymmetries  may indeed give additional gauge or Weyl transformations terms.  

On the Cotton tensor of $Z$, we find
\begin{align}
[\delta_{\epsilon_1}, \delta_{\epsilon_2} ] D\indices{^{ij}_{kl}}[Z] 
&= \frac{\alpha_1^2}{(3!)^3} \partial_r D\indices{^{ij}_{kl}}[Z] \left( \bar{\epsilon}_{2A} \Gamma^r \epsilon_1^A\right) \nonumber \\
&\quad - \frac{\alpha_1^2}{(3!)^3} \mathbb{P}_{(2,2)} \partial_r D_{klpn} \varepsilon^{ijpqr} \bar{\epsilon}_{2A} \Gamma\indices{^n_q} \Gamma^0 \epsilon_1^A \nonumber \\
&\quad + \frac{\alpha_1^2}{2(3!)^3} \mathbb{P}_{(2,2)}\varepsilon^{pqrij} \partial_p D_{qrkl} \bar{\epsilon}_{2A} \Gamma^0
\epsilon_1^A \nonumber \\
&\quad - (1\leftrightarrow2) + (\text{terms containing } A^{AB}_{ij}) .
\end{align}
The first term is a spatial translation. The second term vanishes: it follows from the symplectic Majorana reality conditions that $\bar{\epsilon}_{2A} \Gamma\indices{^n_q} \Gamma^0 \epsilon_1^A$ is symmetric in $1$, $2$, see Appendix \ref{app:Gamma6}. (The other terms are antisymmetric under the exchange of $1$ and $2$.) 
Using the equation of motion for $Z$, the curl appearing in the third term becomes a time derivative. The extra terms containing the bosonic two-forms $A^{AB}_{ij}$ can be shown to vanish. We therefore find indeed a space-time translation,
\begin{equation}
[\delta_{\epsilon_1}, \delta_{\epsilon_2} ] D\indices{^{ij}_{kl}}[Z] = v^\mu \partial_\mu D\indices{^{ij}_{kl}}[Z], \quad v^\mu = - \frac{2 \alpha_1^2}{(3!)^3} (\bar{\epsilon}_{1A} \Gamma^\mu \epsilon_2^A).
\end{equation}

We proceed in a similar fashion for the other fields. We collect here a few identities useful for this computation:
\begin{enumerate}
\item For the commutator on fermionic fields, one needs the following Fierz rearrangement identity, valid for two spinors $\epsilon_1$, $\epsilon_2$ of negative chirality and a spinor $\eta$ of positive chirality:
\begin{equation}
(\bar{\epsilon}_1 \eta) \epsilon_2 = \frac{1}{4}\left[ (\bar{\epsilon}_1 \Gamma^0 \epsilon_2) \Gamma^0 \eta - (\bar{\epsilon}_1 \Gamma^i \epsilon_2) \Gamma_i \eta + \frac{1}{2} (\bar{\epsilon}_1 \Gamma_{0ij} \epsilon_2) \Gamma^{0ij} \eta\right] .
\end{equation}
It follows from the completeness relations of gamma matrices, and from the duality relations between rank $r$ and rank $6-r$ antisymmetric products of gamma matrices \cite{Freedman:2012zz}. It is independent from any Majorana condition on the spinors and is therefore also valid when $\epsilon_1$, $\epsilon_2$ and $\eta$ carry free symplectic indices.
\item The Cotton tensor of $\chi_{ij}$ satisfies the massless Dirac equation
\begin{equation}
\Gamma^0 \Gamma^k \partial_k D_{ij} = \dot{D}_{ij}.
\end{equation}
It is a consequence of its equation of motion \eqref{eq:eomchi} and of the $\Gamma$-tracelessness identity $\Gamma^i D_{ij} = 0$.
\item The scalar field satisfies
\begin{equation}
2\phi^{[ABC|E|} \Omega^{D]N} \Omega_{EM} + 3 \Omega^{[AB} \phi^{CD]NE} \Omega_{EM} + 2 \phi^{[ABC|N|} \delta^{D]}_M = \frac{1}{2} \phi^{ABCD} \delta^N_M .
\end{equation}
This identity follows from
\begin{equation}\label{eq:9indices}
\Omega^{[AA'}\Omega^{BB'}\Omega^{CC'}\Omega^{DD'}\delta^{N]}_M = 0
\end{equation}
by contracting with $\phi^{PQRS}\Omega_{PA'}\Omega_{QB'}\Omega_{RC'}\Omega_{SD'}$. Equation \eqref{eq:9indices} holds because the indices go from $1$ to $8$ and the antisymmetrization is performed on nine indices.
\end{enumerate}
After this is done, we find \begin{equation}
[\delta_{\epsilon_1}, \delta_{\epsilon_2} ] \Phi = - \kappa^2 (\bar{\epsilon}_{1A} \Gamma^\mu \epsilon_2^A) \partial_\mu \Phi
\end{equation}
on any (gauge-invariant) field $\Phi$, provided the following relations hold between the constants $\alpha_i$:
\begin{equation}
\frac{2\alpha_1^2}{(3!)^3} = \alpha_2^2 = \frac{2\alpha_3^2}{3} = \frac{\alpha_4^2}{2} \equiv \kappa^2 . \label{eq:RelAlpha}
\end{equation}
Writing the generetor of supersymmetry transformations as $\bar{\epsilon}_A Q^A$, we therefore get the algebra
\begin{equation}
\{ Q^A_\alpha, Q^B_\beta \} = \kappa^2 \Omega^{AB} \left( P_L \Gamma^\mu C^{-1} \right)_{\alpha\beta} P_\mu . 
\end{equation}

The remaining relations involving the supercharge are easy to derive.    First, one notes that the supersymmetry transformations commute with $\partial_\mu$ since they do not depend explicitly on the spacetime coordinates. So one has  $[Q^A_\alpha, P_\mu] = 0$.  Second, one observes that the established Poincar\'e invariance of the action \cite{Henneaux:1988gg,Henneaux:2016opm} forces the supercharges to transform in a definite representation of the Lorentz group.  Knowing the transformation properties of the supercharges under spatial rotations -- which we do since $SO(5)$ covariace is manifest -- determines the commutators $[Q^A_\alpha, M_{ij}] $ of the supercharges with the spatial rotation generators $M_{ij}$.  The ambiguity as to which representation of the Lorentz group ($(2,1)$ or $(1,2)$) actually occurs is resolved by considering the manifestly Lorentz invariant transformation rules of the scalars, which shows that the supercharges transform in the  $(2,1)$.  Taking into account the $USp(8)$ transformation properties of the supercharges, one thus gets $Q_{\frac12} \sim (2,1; 8)$.

\section{Exotic theories with lower supersymmetry}
\label{sec:lower}

There exist other supersymmetric multiplets containing the exotic graviton, with lower supersymmetry.  These are listed in \cite{deWit:2002vz}.  The corresponding actions are easy to write down by appropriate truncations of the $(4,0)$-theory.  

\subsection{The $(1,0)$-theory}
Consider the $USp(2)$ subgroup of $USp(8)$ acting in the internal $(1,2)$-plane.  Let $a=1,2$ and $\alpha = 3, 4, 5,6,7,8$, so that $(A) = (a, \alpha)$.  We set $A^{12}_{ij} \equiv   A_{ij} = A_{ij} \Omega^{12} $. The conditions
\be
\chi^\alpha_{ij} = 0, \; \; \;  \psi^{ABC} = 0, \; \; \; \phi^{ABCD}= 0, \; \; \; \pi^{ABCD} = 0   \label{eq:Trunc10a}
\ee
and
\be
A^{a \alpha}_{ij} = 0, \; \; \; A^{\alpha \beta }_{ij} = - \frac13 \Omega^{\alpha \beta} A_{ij} \label{eq:Trunc10b}
\ee
are evidently $USp(2)$-invariant and fulfill the $USp(8)$-constraint (\ref{eq:Cons2F}) on the fields.  These conditions are furthermore invariant under the above supersymmetry transformations when the supersymmetry parameters are taken as $(\epsilon^A) = (\epsilon^a,\epsilon^\alpha = 0)$ with $\epsilon^a$ arbitrary, i.e., under  $(1,0)$-supersymmetry.

The conditions \eqref{eq:Trunc10a} and \eqref{eq:Trunc10b} define the $(1,0)$-theory, which contains thus one exotic graviton, one bosonic chiral two-form and two exotic gravitini, which are respectively singlets and doublet of $USp(2)$, corresponding to  
\be (5,1;1) \oplus (3,1; 1)  \oplus (4,1; 2) \, .
\ee
The fields are  $Z_{ijkl}$, $\chi^a_{ij}$ and $A^{ab}_{ij}$, where $a,b$ take two values, $a, b = 1,2$ and the  two-form $A^{12}_{ij}$ is not subject to the condition $\Omega_{ab} A^{ab}_{ij} =0$ (which would eliminate it).  The reality condition $A^*_{ab\, ij} = \Omega_{aa'} \Omega_{bb'} A^{a'b'}_{ij}$ (with $\Omega_{aa'} = \varepsilon_{aa'}$) implies that the singlet two-form $A^{12}_{ij}$ is real.  The action is
\be
S = S_{{\tiny  \yng(2,2)}} +  S_{{\tiny  \yng(1,1)}_F} + 2 S_{{\tiny  \yng(1,1)}_B} .
\ee

\subsection{The $(2,0)$ and $(3,0)$-theories}

The analysis proceeds in the same way for the $(2,0)$ and $(3,0)$-theories.

One finds:
\begin{itemize}
\item For the $(2,0)$-theory, with $R$-symmetry $USp(4)$, one splits the $USp(8)$-indices as $(A)= (a, \alpha)$, $a=1,2,3,4$, $\alpha = 5,6,7,8$.   The truncation to the $(2,0)$-theory is obtained by imposing
\be
\chi^\alpha_{ij} = 0, \; \; \;  \psi^{ab\alpha} = 0, \; \; \; \psi^{a\alpha \beta} = \frac14 \Omega^{\alpha \beta} \psi^{abc} \Omega_{bc} ,\; \; \; \psi^{\alpha \beta \gamma} = 0,  \label{eq:Trunc20a}
\ee
\be
A^{a \alpha}_{ij} = 0, \; \; \; A^{\alpha \beta }_{ij} = \frac14 \Omega^{\alpha \beta} A_{ij}^{ab} \Omega_{ab}, \label{eq:Trunc20b}
\ee
\be
\phi^{abc \alpha}= 0, \; \; \;  \phi^{ab \alpha \beta} = \frac12 \Omega^{\alpha \beta} \Omega^{ab} \phi, \; \; \; \phi^{a \alpha \beta \gamma}= 0, \; \; \;  \phi^{ \alpha \beta \gamma \delta} =   \varepsilon^{\alpha \beta \gamma \delta} \phi
\ee
and
\be
\pi^{abc \alpha}= 0, \; \; \;  \pi^{ab \alpha \beta} = \frac12 \Omega^{\alpha \beta} \Omega^{ab} \pi, \; \; \; \pi^{a \alpha \beta \gamma}= 0, \; \; \;  \pi^{ \alpha \beta \gamma \delta} =   \varepsilon^{\alpha \beta \gamma \delta} \pi .
\ee
The independent fields are $Z_{ijkl}$, $\chi^a_{ij}$, $A^{ab}_{ij}$, $\psi^{abc}$, $\phi^{abcd} = \phi \varepsilon^{abcd}$ and $\pi^{abcd}= \pi \varepsilon^{abcd}$ and are not constrained by $USp(4)$ conditions.  Except for the chiral two-forms, which transform in the ${\mathbf 5} \oplus {\mathbf 1}$, they are all in irreducible representations of $USp(4)$.

The $(2,0)$-theory contains therefore one exotic graviton,  four exotic gravitini, six bosonic chiral two-forms, four spin-$\frac12$ fields and one scalar, 
\be (5,1;1) \oplus (3,1; 1)  \oplus (3,1; 5) \oplus (1,1;1) \oplus (4,1; 4) \oplus (2,1; 4) \, .
\ee
The action is obtained by mere substitution of the truncation conditions into the $(4,0)$-action. 
It is invariant under the supersymmetry transformations that preserve the truncation, i.e., the supersymmetry transformations with parameters taken as $(\epsilon^A) = (\epsilon^a,\epsilon^\alpha = 0)$ where $\epsilon^a$ is arbitrary.   This is $(2,0)$-supersymmetry.

\item For the $(3,0)$-theory, with $R$-symmetry $USp(6)$, one splits the $USp(8)$-indices as $(A)= (a, \alpha)$, $a=1,2,3,4, 5, 6$, $\alpha = 7,8$.  The truncation to the $(3,0)$-theory is obtained by imposing
\be
\chi^\alpha_{ij} = 0, \; \; \;  \psi^{ab\alpha} = 0, \; \; \; \psi^{a\alpha \beta} = \frac12 \Omega^{\alpha \beta} \psi^{abc} \Omega_{bc} ,\; \; \; \psi^{\alpha \beta \gamma} = 0,  \label{eq:Trunc30a}
\ee
\be
A^{a \alpha}_{ij} = 0, \; \; \; A^{\alpha \beta }_{ij} = \frac12 \Omega^{\alpha \beta} A_{ij}^{ab} \Omega_{ab}, \label{eq:Trunc30b}
\ee
\be
\phi^{abc \alpha}= 0, \; \; \;  \phi^{ab \alpha \beta} = \frac12 \Omega^{\alpha \beta}  \phi^{abcd}\Omega_{cd}
\ee
and
\be
\pi^{abc \alpha}= 0, \; \; \;  \pi^{ab \alpha \beta} = \frac12 \Omega^{\alpha \beta}  \pi^{abcd}\Omega_{cd} .
\ee
The independent fields are $Z_{ijkl}$, $\chi^a_{ij}$, $A^{ab}_{ij}$, $\psi^{abc}$, $\phi^{abcd} $ and $\pi^{abcd}$ and  are constrained by the sole $USp(6)$ conditions
\be
\phi^{abcd} \Omega_{ab} \Omega_{cd} = 0, \; \; \;  \pi^{abcd} \Omega_{ab} \Omega_{cd} = 0
\ee
which are consequences of the $USp(8)$ constraints.   Except for the chiral two-forms, which transform in the $\mathbf{14} \oplus {\mathbf 1}$, and the spin-$\frac12$ fields, which transform in the $\mathbf{14}' \oplus {\mathbf 6}$, they are all in irreducible representations of $USp(6)$\footnote{The $\mathbf{14}$ and $\mathbf{14}'$ of $USp(6)$ are the antisymmetric symplectic traceless tensor representations of respective ranks $2$ and $3$.}.

The $(3,0)$-theory with $R$-symmetry $USp(6)$ contains therefore one exotic graviton, six exotic gravitini, fifteen bosonic chiral two-forms, twenty spin-$\frac12$ fields and fourteen scalars, 
 \be (5,1;1) \oplus (3,1; 1)  \oplus (3,1; 14) \oplus (1,1;14) \oplus (4,1; 6) \oplus (2,1; 6) \oplus (2,1; 14') \, .
\ee
The action is obtained by mere substitution of the truncation conditions into the $(4,0)$-action. 
It is invariant under the supersymmetry transformations that preserve the truncation, i.e., supersymmetry transformations with parameters $(\epsilon^A) = (\epsilon^a,\epsilon^\alpha = 0)$ where $\epsilon^a$ is arbitrary.   This is $(3,0)$-supersymmetry. 

\end{itemize}

\section{Dimensional reduction}
\label{sec:DimRed}

The dimensional reduction of the exotic $(4,0)$-supergravity was already discussed in \cite{Hull:2000zn,Hull:2000rr} at the level of the equations of motion.  

The dimensional reduction to 5 spacetime dimensions of the action for a chiral tensor of $\tiny{\yng(2,2)}$-type has been performed in \cite{Henneaux:2016opm}, where it was shown that the prepotential of the chiral tensor correctly yields the prepotential for a single spin-$2$ field, and that the dimensionally reduced action of the chiral tensor becomes the (prepotential version of the) Pauli-Fierz action \cite{Bunster:2013oaa}.

Similarly, the dimensional reduction to 5 spacetime dimensions of the action for a chiral bosonic two-form leads to the action for a single $U(1)$ vector field described by the Maxwell action -- or its dual description in terms of a single two-form --, while the  dimensional reduction of the scalar fields just yields an equal number of scalar fields.

The little algebra for a massless field in 5 dimensions is $so(3)$, while it is $so(4) \simeq so(3) \oplus so(3)$ in 6 dimensions.  For the chiral (anti-chiral) fields only the first (second) $so(3)$ factor acts non trivially; the other factor is represented trivially.  The representation of the little algebra in 5 dimensions inherited from 6 dimensions by dimensional reduction is just the representation of the chiral $so(3)$ non-trivially represented.

The same situation of course prevails for the fermions.  The dimensional reduction of the spin-$\frac12$ fields presents no difficulty, so we focus here only on the remaining fields, the exotic gravitini.  We shall perform the discussion in terms of the spinorial two-form $\Psi_{\lambda \mu}$.  A complete discussion in terms of the prepotentials will be given elsewhere \cite{Future3}.

An explicit realization of the six-dimensional gamma matrices is given by the block form
\begin{align}
\Gamma_\mu &= \sigma_1 \otimes \gamma_\mu = \begin{pmatrix}
0 & \gamma_\mu \\
\gamma_\mu & 0
\end{pmatrix} \quad(\mu = 0, \dots, 4), \\
\Gamma_5 &= \sigma_2 \otimes I= \begin{pmatrix}
0 & - iI \\ iI & 0
\end{pmatrix},
\end{align}
each block being $4 \times 4$. The first five are given in terms of the five-dimensional gamma matrices (in particular  $\gamma_4 = i \gamma_0 \gamma_1 \gamma_2 \gamma_3$ is what is usually called $\gamma_5$). In this representation, the $\Gamma_7$ matrix is diagonal,
\begin{equation}
\Gamma_7 = \begin{pmatrix}
I & 0 \\ 0 & -I
\end{pmatrix} .
\end{equation}
Therefore, a left-handed $2$-form $\Psi_{\mu\nu} = P_L \Psi_{\mu\nu}$ takes the simple form
\begin{equation}
\Psi_{\mu\nu} = \begin{pmatrix}
\hat{\psi}_{\mu\nu} \\ 0
\end{pmatrix}
\end{equation}
and its Dirac conjugate is $\bar{\Psi}_{\mu\nu} = \begin{pmatrix}
0 & \bar{\hat{\psi}}_{\mu\nu}
\end{pmatrix}$ .
The six-dimensional field $\hat{\psi}_{\mu\nu}$ ($\mu,\nu = 0, \dots 5$) reduces to two fields $\psi_{\mu\nu} = \hat{\psi}_{\mu\nu}$ and $\psi_\mu = \hat{\psi}_{\mu5}$ ($\mu,\nu = 0, \dots 4$) in five dimensions.

Using $\Gamma^{\mu_1 \dots \mu_5} = - \varepsilon^{\mu_1 \dots \mu_5 \nu} \Gamma_\nu \Gamma_7$ and $\Gamma_7 \Psi_{\mu\nu} = \Psi_{\mu\nu}$, the Lagrangian for the spinorial $2$-form in six dimensions can be rewritten as
\begin{equation}
\mathcal{L} = - \bar{\Psi}_{\mu\nu} \varepsilon^{\mu\nu\rho\sigma\tau\lambda} \Gamma_\lambda \partial_\rho \Psi_{\sigma\tau} .
\end{equation}
Using the above decomposition, it becomes
\begin{align}
\mathcal{L} = - i \bar{\psi}_{\mu\nu} \varepsilon^{\mu\nu\rho\sigma\tau} \partial_\rho \psi_{\sigma\tau} + 2 \bar{\psi}_{\mu\nu} \varepsilon^{\mu\nu\rho\sigma\tau} \gamma_\rho \partial_\sigma \psi_{\tau} + 2 \bar{\psi}_{\mu} \varepsilon^{\mu\nu\rho\sigma\tau} \gamma_\nu \partial_\rho \psi_{\sigma\tau} .
\end{align}
Now, one can eliminate the five-dimensional spinorial two-form $\psi_{\mu\nu}$ using its own equation of motion.  Indeed, varying the action with respect to $\psi_{\mu\nu}$ yields \begin{equation}
\varepsilon^{\mu\nu\rho\sigma\tau} \partial_\rho \left( i \psi_{\sigma\tau} + 2 \gamma_{\sigma} \psi_\tau \right) = 0
\end{equation}
from which one derives
\begin{equation}
\psi_{\mu\nu} = 2i\gamma_{[\mu} \psi_{\nu]} + \partial_{[\mu} \Lambda_{\nu]}
\end{equation} for some $\Lambda_\nu$. Inserting this expression in the Lagrangian, one gets
\begin{equation}
\mathcal{L} = - 4 i \bar{\psi}_{\mu } \varepsilon^{\mu\nu\rho\sigma\tau} \gamma_{\sigma\tau} \partial_\nu \psi_{\rho }  = - 8 \bar{\psi}_{\mu} \gamma^{\mu\nu\rho} \partial_\nu \psi_{\rho}
\end{equation}
which is exactly the Rarita-Schwinger action for $\psi_\mu$  (after rescaling $\psi_\mu \rightarrow \psi_\mu / 2 \sqrt{2}$)\footnote{Note that the sign of the Rarita-Schwinger action is ``correct'', in the sense that the physical spin-$\frac32$ components and the spin-$\frac12$ field have the same sign for the kinetic term.}.  We can conclude that the dimensional reduction of the chiral spinorial two-form in six dimensions gives correctly a single Rarita-Schwinger field in five dimensions.

If one imposes symplectic Majorana conditions in six dimensions, these simply go through to five dimensions. 

Collecting the individual pieces,  one can now compare the action and supersymmetry transformations of the theory obtained by dimensional reduction of the exotic $(4,0)$-theory with five-dimensional maximal (linearized) supergravity \cite{Cremmer:1979uq}.  It is straightforward to check that they coincide.  Similarly, the exotic $(1,0)$, $(2,0)$ and $(3,0)$-theories yield the (linearized) versions of the $N=2$, $N=4$ and $N=6$ theories in 5 dimensions with $USp(N)$ $R$-symmetry \cite{Cremmer,Chamseddine:1980sp,Awada:1985ep,Gunaydin:1985cu}.

It is interesting to point out that if instead of the standard description of gravity based on a symmetric tensor,   one uses the description involving the dual graviton given by a $\tiny{\yng(2,1)}$-tensor in five dimensions, and keeps the two-form $\psi_{\mu \nu}$ instead of the Rarita-Schwinger field $\psi_\mu$, one gets the dual description of five-dimensional linearized supergravity  alluded to in \cite{Curtright:1980yk}.  [The prepotential formulation of linearized gravity enables one to easily trade $\tiny{\yng(2)}$ for $\tiny{\yng(2,1)}$ and vice-versa \cite{Bunster:2013oaa}.]

\section{Conclusions}
\label{sec:Conclu}

The $(4,0)$ exotic supergravity theory (and its $(1,0)$, $(2,0)$ and $(3,0)$ truncations), which is intrinsically chiral, possesses an action principle which we have explicitly constructed in the free case.  This action principle has unconventional features.  The basic variables of the variational principle are ``prepotentials'' adapted to the self-duality properties of the fields.  Although Poincar\'e invariant, the action is not manifestly so.   We have shown that it has $(4,0)$ supersymmetry and verified the supersymmetry algebra.

The same tension between manifest Poincar\'e invariance and manifest duality symmetry observed in all duality-symmetric formulations (without auxiliary fields)  \cite{Deser:1976iy,Henneaux:1988gg,Schwarz:1993vs,Bunster:2011qp} appears again here.  Perhaps this is a signal that duality symmetry might be more fundamental \cite{Bunster:2012hm}.

The lack of manifest Poincar\'e invariance makes the coupling to gravity more difficult to handle and one must resort to other techniques for that purpose \cite{Henneaux:1988gg}. The fields do not transform with the usual tensorial properties, as also discussed recently in \cite{Sen:2015nph} from a different perspective in a similar context.  How serious this is a drawback in the present case is not clear, however, since exotic supergravities are supposed to contain gravity and for that reason do not need to be coupled to a metric tensor.  Finding the appropriate ``exotic geometry''  remains a challenge.

To become a true physical model, the $(4,0)$-theory should  in any case admit consistent interactions.  The introduction of interactions is notoriously complicated for a collection of chiral two-forms and is expected to be even more so here since there are exotic chiral fields of more intricate types.  An intriguing feature of the prepotentials that enter the description of these exotic fields is that they enjoy a generalized form of Weyl gauge symmetry, which is controlled by appropriate Cotton tensors.  Perhaps the non linear extensions of these Cotton tensors (see  \cite{Linander:2016brv} for nonlinear Cotton tensors in three dimensions) might play an important role in the construction of consistent interactions.

\section*{Acknowledgments} 
V. L. is Research Fellows at the Belgian F.R.S.-FNRS. This work was partially supported by the ERC Advanced Grant ``High-Spin-Grav'', by FNRS-Belgium (convention FRFC PDR T.1025.14 and  convention IISN 4.4503.15) and by the ``Communaut\'e Fran\c{c}aise de Belgique'' through the ARC program.

\begin{appendix}

\section{Spinors and gamma matrices in six spacetime dimensions}
\label{app:Gamma6}

(For more information, see e.g. \cite{Scherk:1978fh,Kugo:1982bn,Freedman:2012zz}).

\vspace{.2cm}

Gamma matrices in six spacetime dimensions are denoted by a Greek capital $\Gamma$ letter.  They satisfy
\be
\Gamma^\mu \Gamma^\nu + \Gamma^\nu \Gamma^\mu= 2 \eta^{\mu \nu}
\ee
where the metric has ``mostly plus'' signature, $ \eta = $ diag$(-+++++)$.  The spatial $\Gamma$-matrices are hermitian while $\Gamma^0$ is anti-hermitian, so that
\be
\left(\Gamma^\mu\right)^\dagger = \Gamma^0 \Gamma^\mu \Gamma^0
\ee
The matrix $\Gamma_7$ is defined as the product of all $\Gamma$-matrices
\be
\Gamma_7 = \Gamma_0 \Gamma_1 \Gamma_2 \Gamma_3 \Gamma_4 \Gamma_5 .
\ee
It anticommutes with the other $\Gamma$-matrices, $\{ \Gamma_7, \Gamma^{\mu} \} = 0$, it is hermitian, $\Gamma_7^\dagger = \Gamma_7$, and it squares to the identity, $\Gamma_7^2 = I$.  Therefore, the chiral projectors
\be
P_L = \frac12 \left( I + \Gamma_7 \right), \; \; \; P_R = \frac12 \left( I - \Gamma_7 \right)
\ee
commute with the Lorentz generators.  One can thus impose the chirality
conditions   $ \psi = P_L  \psi$  ($\Leftrightarrow \Gamma_7 \psi = \psi$) or   $ \psi = P_R  \psi$  ($\Leftrightarrow \Gamma_7 \psi = - \psi$) on the spinors, which are then called (positive chirality or negative chirality) Weyl spinors.

The charge conjugation matrix $C$ is defined by the property
\be
- \left(\Gamma^\mu \right)^T =  C \Gamma^\mu C^{-1}
\ee
It is symmetric and unitary.  The matrices $C\Gamma^{\mu_1 \dots \mu_k}$ are symmetric for $k = 0, 3 \pmod{4}$ and antisymmetric for $k = 1, 2 \pmod{4}$.

Defining also
\be
\mathcal{B} = - i C \Gamma^0
\ee
we have for the complex conjugate $\Gamma$-matrices
\be
\left(\Gamma^\mu \right)^* =  \mathcal{B} \Gamma^\mu \mathcal{B}^{-1}.
\ee
The matrix $ \mathcal{B}$ is unitary. It is antisymmetric and fulfills
\be 
 \mathcal{B}^*  \mathcal{B} = -I \, . \label{BSquare}
 \ee
 One has also
 \be
\left(\Gamma_7 \right)^* =  \mathcal{B} \Gamma_7 \mathcal{B}^{-1}. \label{RealGamma7}
\ee

The complex conjugate spinor $ \psi^*$ transforms in the same way as $\mathcal{B} \psi$, or what is the same,  $ \mathcal{B}^{-1} \psi^*$ transforms in the same way as $\psi$. Using (\ref{RealGamma7}), one furthermore sees that if  $\psi$ is a Weyl spinor of definite (positive or negative) chirality, then $ \mathcal{B}^{-1} \psi^*$ is a Weyl spinor of same chirality.  The positive (respectively, negative) helicity representation is equivalent to its complex conjugate.
 It would be tempting to impose the reality condition $\psi^* = \mathcal{B} \psi$, but this is not possible: the consistency condition $\psi^{**} = \psi$ would impose $\mathcal{B}^*  \mathcal{B} = I$, but this contradicts (\ref{BSquare}). 
 
If we have
several spinors  $\psi^A$ ($A = 1, \cdots, 2n$), however, we can impose the condition
\be
\left( \psi^A \right)^* \equiv \psi_A^* = \Omega_{AB} \mathcal{B} \psi^B  \label{eq:SymMaj}
\ee
where the antisymmetric matrix $\Omega$ is the $2n \times 2n$ symplectic matrix
\be
\Omega = \begin{pmatrix}
0 & 1 & 0 & 0 &  \\
-1 & 0 & 0 & 0 & \cdots \\
0 & 0 & 0 & 1 & \\
0 & 0 & -1 & 0 & \\
& & \vdots &  & \ddots
\end{pmatrix} .
\ee
[For why the internal index is lowered as one takes the complex conjugate, see Appendix \ref{app:usp8}.] Because the real matrix $\Omega$ squares to $-I$, $\Omega^2 = - I$,  the equation (\ref{eq:SymMaj}) consistently implies $\psi^{A**} = \psi^A$.  Spinors fulfilling (\ref{eq:SymMaj})  are called ``symplectic Majorana spinors''.  One can furthermore assume that the $\psi^A$'s are of definite chirality.   The Weyl and symplectic Majorana conditions define together  ``symplectic Majorana-Weyl spinors'' (of positive or negative chirality). For later purposes, we define $\Omega^{AB}$ (with indices up) through $\Omega^{AB}\Omega_{CB} = \delta^A_C$. The matrix $\Omega^{AB}$  is numerically equal to $ \Omega_{AB}$. 

The Dirac conjugate is defined as
\be
\bar{\psi} = i \psi^\dagger \Gamma^0.
\ee
 For symplectic Majorana spinors, it can be written as
 \be
 \bar{\psi}_A = \Omega_{AB} \left(\psi^B\right)^T C.
 \ee
 
 If $\psi^A$  and $\chi^A$ are two symplectic (anticommuting) Majorana spinors, then the product  $ \bar{\psi}_A \chi^A$ is a
real Lorentz scalar, which is symmetric for the exchange of $\psi^A$ with $\chi^A$,
\be
\bar{\psi}_A \chi^A = \bar{\chi}_A \psi^A \, .
\ee
More generally, the products $\bar{\psi}_A \Gamma^{\mu_1 \cdots \mu_k} \chi^A$ are real Lorentz tensors, which are symmetric under the exchange of $\psi^A$ with $\chi^A$ for $k = 0, 3 \pmod{4}$ and antisymmetric for $k = 1, 2 \pmod{4}$.
Moreover, if the spinors have definite chirality, some of these products vanish: if $\chi$ and $\psi$ have the same chirality, $\bar{\psi}_A \Gamma^{\mu_1 \cdots \mu_k} \chi^A$ vanishes for even $k$, while if $\chi$ and $\psi$ have opposite chiralities, it vanishes for odd $k$.

When the symplectic indices are not contracted, the rule for flipping the spinors is the following:
\begin{equation}
\bar{\psi}_A \Gamma^{\mu_1 \cdots \mu_k} \chi^B = (-1)^k\, \Omega^{BC} \Omega_{AD} \, \bar{\chi}_C \Gamma^{\mu_k \cdots \mu_1} \psi^D
\end{equation}
(notice the index reversal). This is useful for the computation of the supersymmetry commutators in section \ref{sec:susy}.

 \section{$USp(8)$ and reality conditions}
 \label{app:usp8}
 
 The group $USp(8)$ is defined as the group of $8 \times 8$ complex matrices that are both unitary and symplectic,
\begin{equation}
USp(8) = U(8) \cap Sp(8,\mathbb{C}) .
\end{equation}
Indices $A,B\dots$ range from $1$ to $8$.
Quantities with indices upstairs transform in the fundamental,
\begin{equation}
v^A \rightarrow S\indices{^A_B} v^B, \qquad S \in USp(8).
\end{equation}
Quantities with indices downstairs transform in the contragredient representation (i.e. with the inverse transpose $(S^{-1})^T$),
\begin{equation}
w_A \rightarrow (S^{-1})\indices{^B_A} w_B, \qquad S \in USp(8).
\end{equation}
Therefore, the contraction $w_A v^A$ is $USp(8)$ invariant. Because $USp(8)$ matrices are unitary, $(S^{-1})^T = S^*$, the contragredient representation is actually the complex conjugate representation $w \rightarrow S^* w$. This motivates the notation
\begin{equation}
(v^A)^* \equiv v^*_A
\end{equation}
for the complex conjugates.
The matrices of $USp(8)$ are also symplectic, $S^T \Omega S = \Omega$ (and also $S \Omega S^T = \Omega$). Together with unitarity, this implies the property
\begin{equation}
\Omega S = S^* \Omega .
\end{equation}
Therefore, the quantity
\begin{equation}
w_A \equiv \Omega_{AB} v^B
\end{equation}
transforms indeed as its indices suggest, i.e. $w \rightarrow S^* w$ when $v \rightarrow Sv$.
Because both $v^*_A$ and $\Omega_{AB} v^B$ transform as quantities with indices down, the contractions $v^*_A v^A$ and $v^A \Omega_{AB} v^B$ are invariant.

Quantities with multiple indices transform in the corresponding tensor product of representations.

\section{Conformal geometry for a spinorial $2$-form in five spatial dimensions}
\label{app:cotton5}

The prepotential for a spinorial two-form in six spacetime dimensions is a spatial object, ``living'' therefore in five spatial dimensions. It is an antisymmetric tensor-spinor $\chi_{ij}$, with gauge symmetries
\begin{equation} \label{eq:gaugeweyl6d}
\delta \chi_{ij} = \partial_{[i} \eta_{j]} + \Gamma_{[i} \rho_{j]} .
\end{equation}
The goal of this section is to construct its ``geometry'', i.e., the invariants that can be built out of $\chi_{ij}$ and its derivatives.  We also derive the main properties of these invariants.  We shall develop the formalism without imposing the chirality condition $\Gamma_7  \chi_{ij} =  \chi_{ij}$.  It can of course be imposed.  In that case, $\Gamma_7$ should be replaced in the formulas below by the identity when acting on spinors.

In line with the terminology used in the general study of higher spin conformal geometry, for which the construction follows exactly the same pattern\footnote{This pattern will be precisely detailed for general half-integer spins in three dimensions in \cite{Future2}.  The pattern itself does not depend on the dimension - although the actual details do.}, we call the $\eta$ transformations ``generalized diffeomorphisms'' and the $\rho$ transformations ``generalized Weyl transformations''.   We shall also deliberately use the names ``Einstein tensor'', ``Schouten tensor'' and ``Cotton tensor'' for the relevant invariant tensors, since this is the appropriate terminology in the higher spin case.

\subsection{Einstein tensor}

First, we construct tensors that are invariant under generalized diffeomorphisms.  Of course, it is just enough to take the exterior derivative of $\chi_{ij}$.  By dualizing, one gets the ``Einstein tensor''
\begin{equation}
G_{ij}[\chi] = \varepsilon_{ijklm} \partial^k \chi^{lm}, \; \; \; G_{ij}[\chi] = G_{[ij]}[\chi]
\end{equation}
(antisymmetric tensor as the field $\chi_{ij}$) which is not only invariant under the $\eta$ transformations, but which is also divergenceless,
\be
\partial_i G^{ij}[\chi] = 0
\ee
(``contracted Bianchi identity'').
We have furthermore the properties
\begin{itemize}
\item $\eta$ triviality criterion:
\begin{equation}
G_{ij}[\chi] = 0 \quad\Leftrightarrow\quad \chi_{ij} = \partial_{[i} \eta_{j]} \;\text{ for some } \eta_i.
\end{equation}
\item Divergence:
\begin{equation}
\partial^i T_{ij} = 0 \quad \Leftrightarrow \quad T_{ij} = G_{ij}[\chi] \;\text{ for some } \chi_{ij} .
\end{equation}
\end{itemize}
These are just two applications of the Poincar\'e lemma, in form degrees $2$ and $3$. The first property implies that the most general invariant under the $\eta$ transformation is a function of the Einstein tensor and its derivatives. 

Under generalized Weyl transformations, we have
\begin{equation}
\delta G_{ij} = \varepsilon_{ijklm} \Gamma^l \partial^k \rho^m = - \Gamma_{ijkm} \Gamma_7 \Gamma_0 \partial^k \rho^m
\end{equation}
which gives for the traces
\begin{align}
\delta ( \Gamma^j G_{ij} ) &= 2 \Gamma_{ikm} \Gamma_7 \Gamma_0 \partial^k \rho^m \;\Rightarrow\; \delta ( \Gamma_{[i} \Gamma^k G_{j]k} ) = 2 \Gamma_{[i} \Gamma_{j]km} \Gamma_7 \Gamma_0 \partial^k \rho^m \\
\delta ( \Gamma^{ij} G_{ij} ) &= 6 \Gamma_{km} \Gamma_7 \Gamma_0 \partial^k \rho^m \;\Rightarrow\; \delta ( \Gamma_{ij} \Gamma^{kl} G_{kl} ) = 6 \Gamma_{ij} \Gamma_{km} \Gamma_7 \Gamma_0 \partial^k \rho^m .
\end{align}

\subsection{Schouten tensor}
\label{app:subSchouten}
We define the Schouten tensor as
\begin{align}
S_{ij} &= G_{ij} + \Gamma_{[i} \Gamma^k G_{j]k} - \frac{1}{6} \Gamma_{ij} \Gamma^{kl} G_{kl}
\end{align}
It is again an antisymmetric tensor, which transforms as
\begin{align}
\delta S_{ij} = \left( - \Gamma\indices{_{ij}^{km}} + 2 \Gamma_{[i} \Gamma\indices{_{j]}^{km}} - \Gamma_{ij} \Gamma^{km}  \right) \Gamma_7 \Gamma_0 \partial_k \rho_m 
\end{align}
under generalized Weyl transformations.
Using the identites $\Gamma_{[m} \Gamma\indices{_{n]}^{rs}} = \Gamma\indices{_{mn}^{rs}} + 2 \delta^{[r}_{[m} \Gamma\indices{^{s]}_{n]}}$  and $\Gamma_{mn} \Gamma^{rs} = \Gamma\indices{_{mn}^{rs}} + 4 \delta_{[m}^{[r} \Gamma\indices{^{s]}_{n]}} -2 \delta^{rs}_{mn}$, it can be seen that the combination in brackets reduces to $2 \delta^{km}_{ij}$, so that the Schouten simply transforms as
\begin{align}
\delta S_{ij} = \partial_{[i} \nu_{j]}, \quad \nu_j = 2 \Gamma_7 \Gamma_0 \rho_j .
\end{align}
It is this simple transformation law of the Schouten tensor that motivates its definition.

Using $\Gamma_i \Gamma^k = \delta^k_i + \Gamma\indices{_i^k}$, the Schouten tensor can be rewritten as
\begin{equation}
S_{ij} = - \left( \delta^{[k}_{[i} \Gamma\indices{_{j]}^{l]}} + \frac{1}{6} \Gamma_{ij} \Gamma^{kl} \right) G_{kl}
\end{equation}
and then, using the identity $\delta^{rs}_{mn} = \frac12 \left(  \delta^{[p}_{[m} \Gamma\indices{_{n]}^{q]}} + \frac16 \Gamma_{mn} \Gamma^{pq} \right) \Gamma\indices{_{pq}^{rs}}$, we can write the Einstein tensor in terms of the Schouten tensor as
\begin{equation}
G_{ij} = - \frac{1}{2}\Gamma_{ijkl} S^{kl} . 
\end{equation}
Therefore, the Schouten satisfies
\begin{equation}
\Gamma_{ijkl} \partial^j S^{kl} = 0
\end{equation}
as a consequence of $\partial^j G_{ij} = 0$.
We have also the following direct properties:
\begin{itemize}
\item Pure gauge property:
\begin{equation}
S_{ij}[\chi] = \partial_{[i} \nu_{j]} \quad\Leftrightarrow\quad \chi_{ij} = \partial_{[i} \eta_{j]} + \Gamma_{[i} \rho_{j]} \;\text{ for some } \eta_i \quad (\rho_i = - \frac{1}{2} \Gamma_0 \Gamma_7 \nu_i).
\end{equation}
\item Constraint property:
\begin{equation}
\Gamma_{ijkl} \partial^j T^{kl} = 0 \quad\Leftrightarrow\quad T_{ij} = S_{ij}[\chi] \;\text{ for some } \chi_{ij} .
\end{equation}
This is the property that underlies the introduction of the prepotential for the chiral spinorial two-form $\Psi_{ij}$.
\end{itemize}
Using the identities $\Gamma_{mn} \Gamma^{rs} = \Gamma\indices{_{mn}^{rs}} + 4 \delta_{[m}^{[r} \Gamma\indices{^{s]}_{n]}} -2 \delta^{rs}_{mn}$ and $\Gamma^{ijkl} = - \varepsilon^{ijklm} \Gamma_m \Gamma_0 \Gamma_7$, we can also rewrite the Schouten tensor as
\begin{equation} \label{a49}
S_{ij} = \frac{1}{3} G_{ij} + \frac{1}{3} \Gamma\indices{_{[i}^k} G_{j]k} + \frac{1}{6} \varepsilon_{ijklm} \Gamma^k \Gamma_0 \Gamma_7 G^{lm} .
\end{equation}

\subsection{Cotton tensor}
The Cotton tensor is defined by taking the curl of the Schouten tensor
\begin{equation}
D_{ij}[\chi] = G_{ij} [S[\chi]] = \varepsilon_{ijklm} \partial^k S^{lm}[\chi] .
\end{equation}
It is invariant by construction under both generalized diffeomorphisms and Weyl transformations \eqref{eq:gaugeweyl6d}. It is divergenceless, $\partial_i D^{ij} = 0$, and also gamma-traceless ($\Gamma^i D_{ij} = 0$) because of $\Gamma_{ijkl} \partial^j S^{kl} = 0$.
Its key properties are:
\begin{itemize}
\item Pure gauge condition: the prepotential $\chi_{ij}$ is pure gauge if and only if its Cotton tensor vanishes,
\begin{equation}
D_{ij}[\chi] = 0 \quad\Leftrightarrow\quad \chi_{ij} = \partial_{[i} \eta_{j]} + \Gamma_{[i} \rho_{j]} \;\text{ for some } \eta_i \text{ and } \rho_i.
\end{equation}
This also means that any invariant under (\ref{eq:gaugeweyl6d}) is a function of the Cotton tensor and its derivatives.
\item Tracelessness and divergencelessness conditions: if a spinorial antisymmetric tensor is both divergenceless and $\Gamma$-traceless, then it is equal to the Cotton tensor of some antisymmetric spinorial prepotential,
\begin{equation}
\partial_i T^{ij} = 0, \quad \Gamma^i T_{ij} = 0 \quad\Leftrightarrow\quad T_{ij} = D_{ij}[\chi] \;\text{ for some } \chi_{ij} .
\end{equation}
\end{itemize}
Both results directly follow from the Poincar\'e lemma and the above definitions.

Using formula \eqref{a49} for the Schouten tensor, we can express the Cotton tensor in terms of the Einstein tensor as
\begin{equation}
D_{ij} = \frac{1}{3} \left( \varepsilon_{ijklm} \partial^{k} G^{lm} + \varepsilon_{ijklm} \partial^k \Gamma^{lp} G\indices{^m_p} + 2 \partial^k \Gamma_k \Gamma_0 \Gamma_7 G_{ij} \right) .
\end{equation}
The second term can also be written as $\frac{1}{3} \Gamma^{pq} \varepsilon_{pqkl[i} \partial^k G\indices{^l_{j]}}$ using the identity\begin{equation}
\varepsilon_{abcdi} \Gamma^{ij} = \frac{1}{2} \left( \delta^j_a \varepsilon_{bcdkl} - \delta^j_b \varepsilon_{acdkl} + \delta^j_c \varepsilon_{abdkl} - \delta^j_d \varepsilon_{abckl} \right) \Gamma^{kl}.
\end{equation}
Note also that the identities
\begin{align}
\Gamma^{a_1 \dots a_k i} D_{ij} &= - k \Gamma^{[a_1 \dots a_{k-1} } D\indices{^{a_k]}_j} \\
\Gamma^{a_1 \dots a_k i j} D_{ij} &= - k (k-1) \Gamma^{[a_1 \dots a_{k-2}} D^{a_{k-1} a_k]}
\end{align}
follow from the gamma-tracelessness of the Cotton tensor.

\end{appendix}

\end{document}